\documentclass[%
reprint,
superscriptaddress,
amsmath,amssymb,
floatfix,
]{revtex4-2}
\usepackage[utf8]{inputenc}

\usepackage{amsthm}

\usepackage[normalem]{ulem} % for strikethrough text

\usepackage{natbib}		%% to make references easier
\setcitestyle{square,numbers}
\usepackage{color}

\usepackage[breaklinks=true,%
colorlinks=true,%
linkcolor=blue,%
urlcolor=blue,%
citecolor=blue]{hyperref}

\usepackage{amsmath}
\usepackage{amssymb}

\usepackage{bbold} %nice identity symbol

\newcommand{\bra}[1]{\ensuremath{\left\langle#1\right|}}
\newcommand{\ket}[1]{\ensuremath{\left|#1\right\rangle}}
\newcommand{\braket}[2]{\ensuremath{\left\langle#1\middle\vert#2\right\rangle}}

\newcommand{\ii}{\text{i}}

\newcommand{\pt}{\text{PT}}
\newcommand{\apt}{\text{APT}}
\newcommand{\tr}{\text{Tr}}

\usepackage{algpseudocode}
\usepackage{algorithm}

\usepackage{graphicx}
\usepackage{subfigure}
\graphicspath{ {figs/} } %% tidy subdir for figure files

\usepackage{tikz}
\usetikzlibrary{quantikz}

\begin{document}
	
	%\preprint{} and  Damped Oscillation of non-Hermitian Quantum Rényi Entropy
	
	\title{ Continuous Phase Transition  in Anyonic-PT Symmetric Systems}
	
		\author{Zhihang Liu}
	\affiliation{Department of Physics, College of Science, North China University
		of Technology, Beijing 100144,  China.}
	
\author{Chao Zheng}
\email{czheng@ncut.edu.cn}  
\affiliation{Department of Physics, College of Science, North China University
	of Technology, Beijing 100144, China.}
\affiliation{School of Energy Storage Science and Engineering, North China University of Technology, Beijing 100144, China.}
\affiliation{Beijing Laboratory of New Energy Storage Technology, Beijing 100144, China}

	\date{\today}
	
	\begin{abstract}
		   We reveal the  continuous phase transition   in  anyonic-PT symmetric systems, contrasting with the  discontinuous phase transition corresponding to the discrete (anti-) PT symmetry.   The continuous  phase transition originates from the  continuity of  anyonic-PT symmetry.  We find  there are  three   information-dynamics patterns for  anyonic-PT symmetric systems: damped oscillations with an overall decrease (increase) and asymptotically  stable damped oscillations,   which are three-fold degenerate and distorted using the Hermitian quantum  Rényi entropy  or distinguishability. It is the  normalization of the non-unitary evolved density matrix causes the degeneracy and distortion.  We give a  justification for  non-Hermitian quantum Rényi entropy being negative. By exploring the mathematics and physical meaning of  the negative entropy in open quantum systems, we connect the  negative non-Hermitian quantum Rényi entropy and negative quantum conditional entropy,   opening up a new journey to rigorously investigate the negative entropy in open quantum systems. 
	\end{abstract}
	
	\maketitle

	\section{Introduction}
	
	 The two fundamental discrete symmetries in physics are given by  the parity operator  P and   the time reversal operator T.
	In recent decades, parity-time (PT) symmetry and its spontaneous symmetry breaking  attracts growing interesting both in theory and experiments. On one hand, non-Hermitian (NH) physics with parity-time symmetry can be seen as a complex extension of the conventional quantum mechanics, having novel properties. On the other hand, it closely related to open and dissipative systems of realistic physics \cite{breuer2002theory, barreiro2011open, hu2020quantum, del2020driven, zheng2021universal}.   Symmetries, such as the parity-time-reversal (PT) symmetry \cite{bender1998real, kawabata2017information, zheng2013observation, PhysRevLett.132.083801},  anti-PT (APT) symmetry \cite{yang2017anti, li2019anti, peng2016anti, bergman2021observation, yang2020unconventional, choi2018observation, zheng2019duality, wen2020observation}, pseudo-Hermitian symmetry \cite{mostafazadeh2002pseudo, mostafazadeh2004pseudounitary, jin2022unitary, xu2023pseudo},  anyonic-PT symmetry  \cite{zheng2022quantum, longhi2019anyonic, arwas2022anyonic} play a central role  in typical NH systems.
	In the quantum regime, various aspects of PT symmetry have been studied, such as   Bose-Einstein condensates \cite{klaiman2008visualization, konotop2016nonlinear},  entanglement \cite{chen2014increase, lee2014entanglement, couvreur2017entanglement}, critical phenomena \cite{ashida2017parity, kawabata2017information}, and etc. For a PT-symmetric system, the Hamiltonian $H_{{\rm PT}}$ satisfies $[{\rm PT}, H_{{\rm PT}}]=0$. It is in PT-unbroken phase if each eigenstate of Hamiltonian  is simultaneously the eigenstate of the  PT operator, in which case the entire spectrum is real.  Otherwise, it is in PT symmetry  broken phase, and some pairs of eigenvalues become complex conjugate to each other.   Between the two phases lies  exceptional points (EPs) where an unconventional phase transition  occurs  \cite{goldenfeld2018lectures, bender1998real, ashida2020non, li2019observation, zhang2024localization},   and this is related to many intriguing phenomena \cite{wang2022quantum, wu2014non, lin2011unidirectional, regensburger2012parity, kawabata2017information}.
	
	 Anyonic-PT symmetry can be seen as  the complex generalization of PT symmetry and the relationships between  PT, APT, and anyonic-PT symmetry can  be an analogy to relationships between boson, fermion, and anyon \cite{longhi2019anyonic, zheng2022quantum, arwas2022anyonic}. In this spirit, it was  named anyonic-PT symmetry.
	While (anti-) PT symmetry  are discrete, anyonic-PT symmetry  is continuous respect to the phase parameter in a way similar to rotation symmetry \cite{goldenfeld2018lectures}. The investigation of information dynamics in (anti-) PT symmetric systems \cite{kawabata2017information, wen2020observation} shows that the phase transitions in (anti-) PT symmetric systems  are discontinuous. 
	In this paper, through a new information-dynamics description, which is found to be  synchronous and correlated with NH quantum Rényi entropy \cite{renyi1961measures, li2022non, muller2013quantum}, we investigate the non-Hermitian (NH) quantum Rényi entropy dynamics of anyonic-PT symmetric  systems.  Our results    show: in contrast to the discontinuous phase transition in  (anti-) PT-symmetric  systems,   the phase transition in anyonic-PT symmetry is  continuous, and the continuous  phase transition originates from the interplay of features of (anti-) PT symmetry and  the continuity of  anyonic-PT symmetry.  
	 
	While Hermiticity ensures the conservation of probability in an isolated quantum system and guarantees the real spectrum of eigenvalue of energy, it is ubiquitous in nature that the probability in an open quantum system  effectively becomes non-conserved due to the flows of energy, particles, and information between the system and the external environment \cite{ashida2020non}.   	In the study of  radiative decay in reactive nucleus, which is  analyzed by an effective NH Hamiltonian, the essential idea is that  the decay of the norm of a quantum state  indicates the presence of nonzero probability flow to the outside of nucleus \cite{feshbach1958unified, feshbach1962unified}. The non-conserved norm indicates there is information flow between the NH system and environment. Thus, the non-conserved norm is essential for describing information dynamics in NH systems. 
	In quantum information,   trace of  density matrix is a central concept in various formulae characterizing information   properties, such as  von Neumann entropy \cite{von2018mathematical}, Rényi entropy \cite{renyi1961measures, muller2013quantum, fehr2014conditional,  li2022non} and trace distance measuring the distinguishability of two quantum states \cite{nielsen2010quantum, kawabata2017information, xiao2019observation, wen2020observation}.  %Here comes the question that 

In this work, we investigate the NH quantum Rényi entropy dynamics of anyonic-PT symmetric  systems through a new information-dynamics description, which is found to be  synchronous and correlated with NH quantum Rényi entropy.  Our results show that the intertwining of   (anti-) PT  symmetry  leads to  new   information-dynamics patterns: damped oscillation with an overall decrease (increase) and asymptotically  stable damped oscillation.  The approaches of  Hermitian quantum Rényi  entropy or distinguishability adopted in \cite{kawabata2017information, xiao2019observation, bian2020quantum, wen2020observation}
not only degenerate the three distinguished  patterns to the same one, but also  distort it.
The  degeneracy  is caused by the normalization of the non-unitary evolved density matrix, which  leads to the loss of information  about the total probability flow between the open  system and the environment, while our approach based on the non-normalized density matrix reserves all the information related to the non-unitary  time evolution.  
Furthermore, our results show that the lower bounds of both von Neumann  entropy and  distinguishability being zero is related to their distortion of the information dynamics in  the NH  systems. 
The  discussion of  the degeneracy and distortion  also serves as  a  justification for  NH quantum Rényi entropy being negative. We further  explore the mathematical reason and physical meaning of  the negative entropy in open quantum systems,  revealing a  connection between negative NH entropy and negative quantum conditional entropy as both quantities can be negative for similar mathematical reasons.   
Since the physical interpretation and the following applications   of negative quantum conditional entropy  are  successful and  promising  \cite{horodecki2005partial, rio2011thermodynamic, cerf1997negative}, we remark that  our work opens up the new  journey of rigorously investigating  the  physical interpretations and the application  prospects of negative entropy in open quantum  system. Last but not least,  in contrast to the discontinuous phase transition in  (anti-) PT-symmetric  systems,  we find that   the phase transition in anyonic-PT symmetry is  continuous and the continuous  phase transition originates from the interplay of features of (anti-) PT symmetry and  the continuity of  anyonic-PT symmetry.

%This paper is structured as follows. 
%In section \uppercase\expandafter{\romannumeral2}, for generic PT symmetric systems in finite dimensions with non-degenerate eigenvalues, we discuss the quantum recurrence in PT-unbroken open quantum systems and why the recurrence fails at PT-broken open quantum systems.
%%In section \uppercase\expandafter{\romannumeral2}, we  systematically  investigate the non-conservative probability  of  three kinds of PT-symmetry related systems by analyzing   ${\rm Tr}\,\Omega (t)$.
%In section \uppercase\expandafter{\romannumeral3},   we establish the   information-theoretic characterization of  the PT, APT and anyonic-PT symmetric systems with non-Hermitian  entropy $S_{1}(\Omega)$ and   find that   $- \ln {\rm Tr}\,\Omega (t)$ (negative of  the logarithm of ${\rm Tr}\,\Omega (t)$) and $S_{1}(\Omega)$  are highly correlated.
%In section \uppercase\expandafter{\romannumeral4}, we compare our approach with the approaches of conventional entropy and  trace distance adopted in previous works investigating  non-Hermitian systems. We obtain  similar results for PT symmetric systems, 
%while new patterns of the entropy dynamics associated with the APT or anyonic-PT symmetric Hamiltonians  are revealed by our approach. 

\section{NH Quantum Rényi entropy in anyonic-PT symmetric systems}	

\begin{figure}[t]
	\includegraphics[width=\linewidth]{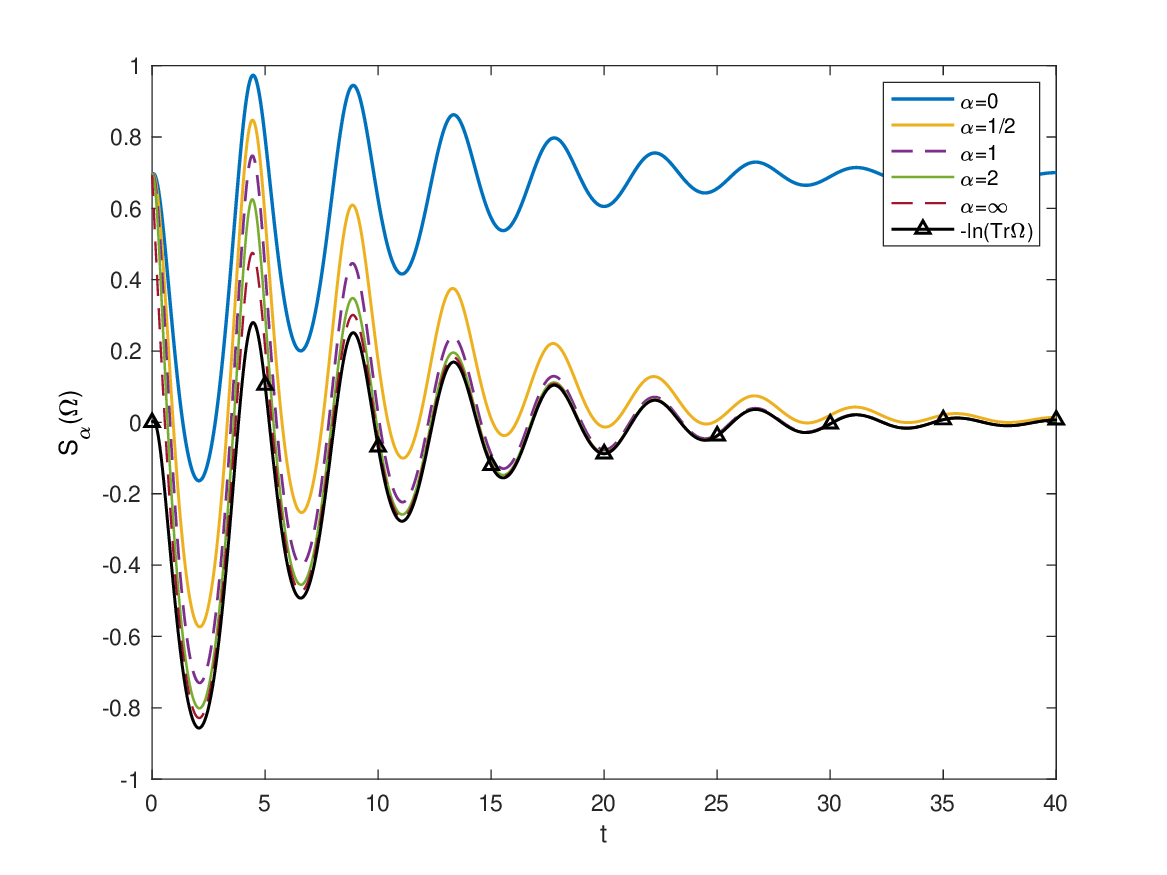}  %fvarphi
	\caption{       $\varphi =- \pi/18$. $\lambda=0$, $\delta>0$, $H_{\varphi}$ of Eq.(\ref{varphi}) in PT-unbroken phase.  $S_{\alpha}(\Omega)$ behave similarly for typical $\alpha$. The black line marked with  $\triangle $  represents $-\ln {\rm Tr }\,\Omega (t)$, which is showed to be synchronous and highly correlated with $S_{\alpha}(\Omega)$.
		\label{alpha}       }
\end{figure}

Quantum Rényi entropy \cite{muller2013quantum} is suited for  Hermitian quantum systems (thus we call it Hermitian quantum Rényi entropy) as it    requires the trace of density matrix  to satisfy  $\tr \,  \rho \in (0,1]$.  %which  depends on incomplete and complete probability distributions.  
The Hermitian quantum Rényi entropy is defined as:
\begin{equation}
	\label{e6}
	S^H_{\alpha}(\rho)=\frac{\ln \tr \, \rho^\alpha}{1-\alpha},
\end{equation}
where $ \alpha\in(0,1)\cup(1,\infty)$. 
%where the trace of density matrix satisfies  $ \tr  \, \rho=1$.  
%For Hermitian systems, $\ln \tr\, \rho $ is always zero, which  means there is no information flow between the systems and environment, which is ,closed , mixed state,  {\'e}
If the initial quantum state $\rho(0)$ is a pure state, $S^H_{\alpha}(\rho)$ is trivial as it is always zero under unitary time evolution.
For open quantum systems with  the trace of initial density matrix less than 1,  due to the nonzero probability flow between the systems and environment, $\tr \, \rho > 1$ is possible  with the time evolution of the  systems.
%the trace of density matrix can be greater than 1 
Thus,  the condition  $\tr \,  \rho \in (0,1]$  should be relaxed to $\tr \,  \rho \ge 0$  for open quantum systems.
To describe the information dynamics in  NH open quantum systems properly,  NH quantum Rényi entropy \cite{li2022non} is defined using both the non-normalized density matrix $\Omega$ and the normalized one $\rho=\Omega/ \tr \, \Omega$ as   
\begin{equation}
	\label{e11}
	S_{\alpha}(\Omega)=
		\frac{\ln \tr (\Omega^{\alpha-1}\rho)}{1-\alpha} \quad  \alpha\in(0,1)\cup(1,\infty), 
\end{equation}
with $ S_{0,1,\infty}(\Omega)=S_{\alpha\rightarrow 0,1,\infty}(\Omega)$, 
\begin{equation}
	S_{1}(\Omega)= -{\rm Tr}\, ( \rho \ln \Omega). \label{s1}
\end{equation}     
Another commonly adopted description of information dynamics is distinguishability $D$ of two quantum states  \cite{breuer2009measure, nielsen2010quantum, kawabata2017information}, 
\begin{gather}
	D \left( \rho_{1} \left( t \right), \rho_{2} \left( t \right) \right)
	= \frac{1}{2} \, \mathrm{Tr} \left|\,\rho_{1} \left( t \right) - \rho_{2} \left( t \right)\,\right| , \label{Dist}
\end{gather}
where $| \rho | := \sqrt{\rho^{\dag} \rho}$,  $\rho_{1, 2}$ are normalized density matrices.
We notice that  the only difference between the expressions of $S_{\alpha}(\Omega)$ and $S^H_{\alpha}(\rho)$ is the use of $\Omega $. 
Investigation of 	$S_{1}(\Omega)$ is enough for our purpose, as the dynamics of NH  quantum Rényi entropy   for different $\alpha$ is similar \cite{li2022non,muller2013quantum}. 
Boltzmann's  entropy formula and Shannon's  entropy formula state  the logarithmic connection between entropy and probability. We borrow this wisdom and take the natural logarithm of $\tr \, \Omega$.  
We   find  $-\ln {\rm Tr }\,\Omega (t)$ can serve as a new description for the information dynamics in NH systems,  as it  is found to be  synchronous and correlated with NH quantum Rényi entropy. $-\ln {\rm Tr }\,\Omega (t)$  captures the essence of the information dynamics in NH systems as we show in FIG.(\ref{alpha}).
%	Thus,  we propose that  $-\ln {\rm Tr }\,\Omega (t)$ can serve as a unification of NH quantum Rényi entropy and brings out the essence of . 	%(trace of density matrix).  (Hermitian)

\subsection*{Anyonic-PT symmetry}

%with $\triangle $ 

	\begin{figure}[t]
	\includegraphics[width=\linewidth]{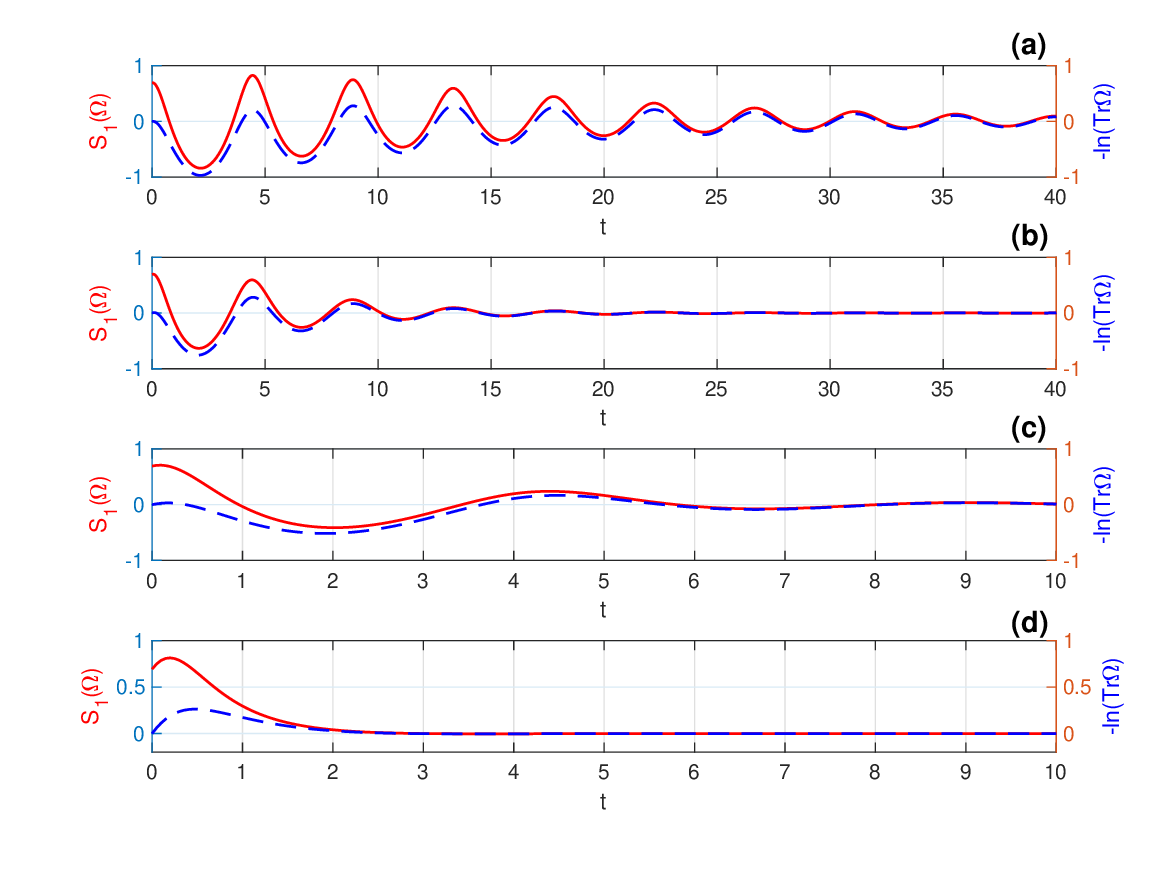}  %fvarphi
	\caption{    % parameter setting:  $r=r_{1}=1$, $\theta_{1} =\pi/6$, $\theta=3\pi/4$;  we then have  $\delta>0$ and  
		The  red line represents $S_{1}(\Omega)$, the dashed blue line represents $-\ln {\rm Tr }\,\Omega (t)$. 
		$\lambda=0$,   $r \cos \theta = - \sqrt{2}/2$,  $S_{1}(\Omega) $  is  asymptotically  stable. $\delta>0$, $H_{\varphi}$ in PT-unbroken phase.
		(\textbf{a})  $\varphi=-\pi/36$,  
		(\textbf{b}) $\varphi=-\pi/12$,  
		(\textbf{c})  $\varphi=-\pi/6$,  
		(\textbf{d})  $\varphi=-3\pi/4$. Clearly, the relaxation time of the  damped oscillation is determined  by  $q \cdot 2r\cos \theta$.
		\label{fvarphi}       }
\end{figure}

\begin{figure}[t]
	\includegraphics[width=\linewidth]{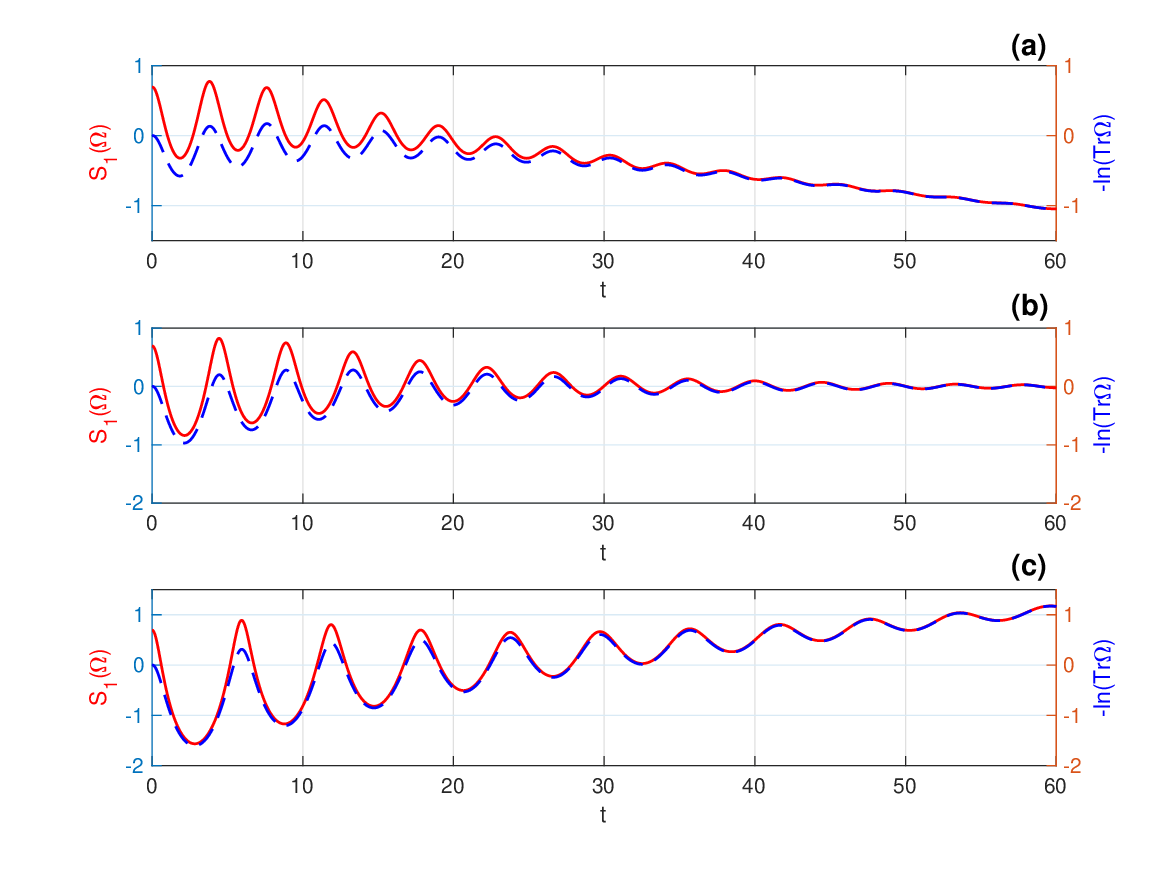} % pheno2
	\caption{    %  parameter setting:  $r_{1}=1$, $\theta_{1} =\pi/6$, ;  $\theta=3\pi/4$,
		$\varphi=-\pi/36$. $\delta>0$, $H_{\varphi}$ in PT-unbroken phase.
		(\textbf{a})  $r=0.8$,   $\lambda>0$; 
		(\textbf{b}) $r=1$,      $\lambda=0$;
		(\textbf{c})  $r=1.2$,  $\lambda<0$. The   three   information-dynamics patterns: damped oscillation with an overall decrease (increase) and asymptotically  stable  damped oscillation are well predicted by   Eq.(\ref{varlim}). 
		%(\textbf{d})  $\varphi=-\pi/48$, $r=1$, $\theta=13\pi/24$, $\lambda=0$.
		\label{phe}       }
\end{figure}

\begin{figure}[t]
	\includegraphics[width=\linewidth]{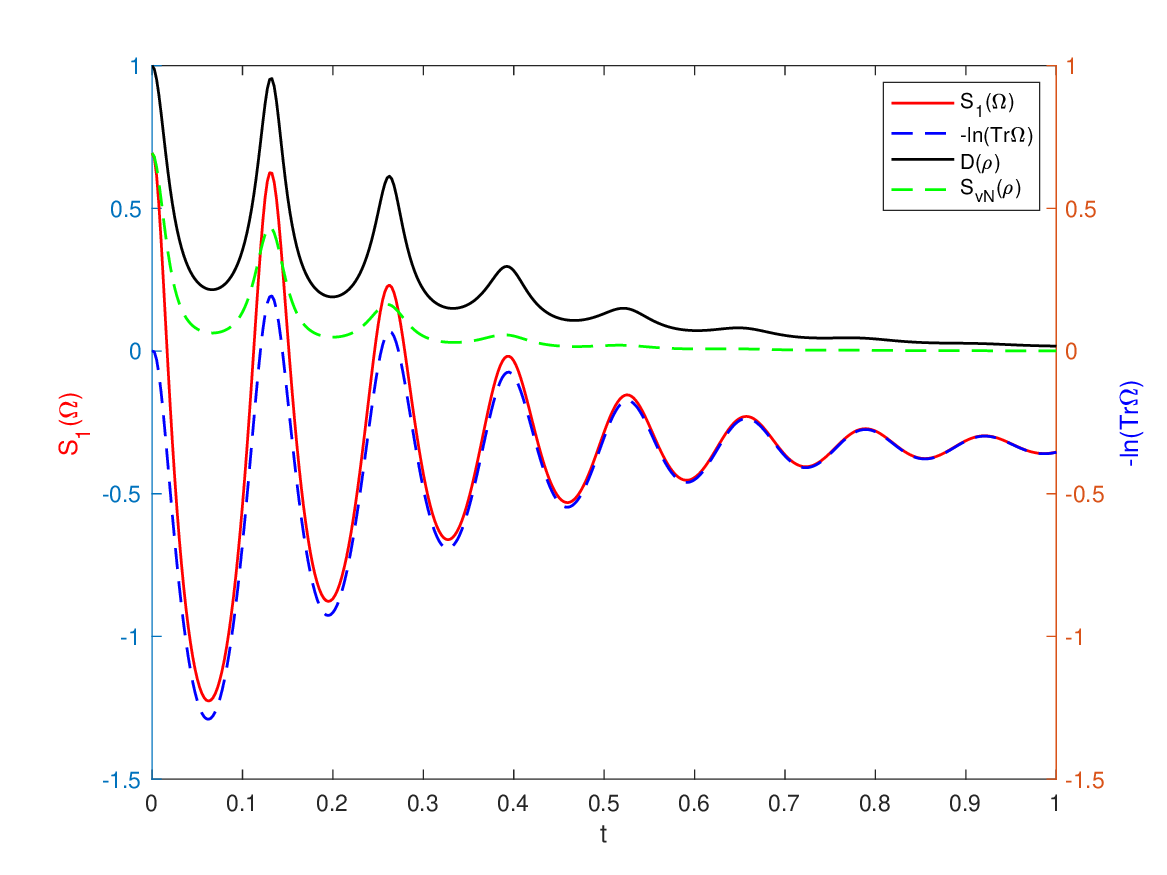} 
	\caption{  $r=40$, $r_{1}=32$, $\theta=33 \pi /64 $,  $\varphi=-2 \arctan \sqrt{\frac{-\delta}{r^{2} \cos^{2} \theta}} $,  $\delta<0$,   $H_{\varphi}$ in PT-broken phase.		\label{continuous}       }
\end{figure}

 Anyonic-PT symmetric Hamiltonians  $H_{\varphi } $ satisfy $({\rm PT})H_{\varphi }({\rm PT})^{-1}=e^{\ii \varphi }H_{\varphi }$, and thus
 \begin{equation}
 	\begin{aligned}
 	H_{\varphi } =e^{-\ii \frac{\varphi }{2} } H_{{\rm PT}} = p H_{{\rm PT}}+q ( \ii H_{{\rm PT}}), 
 	\label{anyon}
 	\end{aligned}
 \end{equation}
 where $p=\cos \frac{\varphi}{2},  q=-\sin \frac{\varphi}{2}$,  $H_{\pt}$ are PT-symmetric and  $\ii H_{{\rm PT}}$ satisfy anti-PT symmetry (thus we denote $H_{{\rm APT}}=\ii H_{{\rm PT}}$).  $H_{\pt}$ and  $H_{\apt}$ commute, which  means  the two can be simultaneously diagonalized and so the eigenfunctions of $H_{\varphi } $ ($H_{\pt}$ and  $H_{\apt}$) are independent of $\varphi$ even though the eigenvalues vary with $\varphi$.  The eigenvalues of $H_{\pt}$ ($H_{\apt}$)   undergo a abrupt change with the symmetry breaking, indicating a discontinuous phase transition \cite{kawabata2017information, wen2020observation}.  In contrast, the change of eigenvalues of $H_{\varphi}$ with the symmetry breaking can be continuous because of  the phase $e^{-\ii \frac{\varphi }{2} }$, indicating the possibility of continuous phase transition.  We reveal the continuous phase transition in anyonic-PT symmetric systems by investigating its information dynamics. 
 
 % So, the information dynamics of $H_{\varphi } $ ($H_{\pt}$ and  $H_{\apt}$) varies with eigenvalues.
 %$H_{\varphi }$ allows us to take a unified treatment to  $H_{{\rm PT}}$ (when $\varphi=0$) and  $H_{{\rm APT}}$  (when $\varphi=-\pi$).  For the general case ($H_{\varphi }\ne H_{{\rm PT} } \; {\rm or}  \; H_{{\rm APT}} $), new novel phenomena and application prospects emerge. 
We employ the usual Hilbert-Schmidt inner product  when  we investigate the effective non-unitary dynamics of  open quantum systems governed by $H_{\varphi } $  \cite{kawabata2017information, brody2016consistency, brody2012mixed},
 \begin{equation}
 	\Omega (t)=e^{-\ii H_{\varphi } t } \, \Omega (0) \, e^{\ii H_{\varphi } ^{\dagger }t}  , \label{dy}
 \end{equation}
 \begin{equation}
 	\rho(t)=\Omega(t)/{\rm Tr}\, \Omega(t) \label{normal}.
 \end{equation}
%For  $H_{\pt}$ in PT-unbroken phase,  
For ${H}_{\varphi}$ with     eigenenergies $E_{n} + \ii \Gamma_{n}$, 
\begin{equation}
	{H}_{\varphi} \ket{\varphi_{n}} = \left( E_{n} + \ii \Gamma_{n} \right) \ket{\varphi_{n}},
\end{equation}
with $  \braket{\varphi_{n} | \varphi_{n} }=1$. 
Define the eigenstates with the largest (second largest) imaginary part as $ \ket{\varphi_{1}}  $ ( $\ket{\varphi_{2}}  $). After a sufficiently long time, $ \ket{\varphi_{1}}  $ and $\ket{\varphi_{2}}  $ dominate the dynamics.  %\cite{kawabata2017information}
%After a sufficiently long time, both $\ket{\varphi_{1}}$ and $\ket{\varphi_{2}}$ dominate the dynamics, and hence the PT dynamics asymptotically behaves as 
With arbitrary initial state $\ket{\varphi_{0}}= { \sum_{n=1}}  c_{n} \ket{\varphi_{n}}  $  and  $\Omega(0)= \ket{\varphi_{0}} \bra{\varphi_{0}}$,
 we have 
\begin{equation}
	\begin{aligned}
		&-\ln  {\rm Tr} \, \Omega(t) \sim   -\ln [\,  |c_{1}| ^{2}  \, e^{2\Gamma_{1}t}+ |c_{2}| ^{2} \,  e^{2\Gamma_{2}t}  +\\ &  (  c_{1} c_{2}^{\dagger} \,  e^{-\ii (E_{1} - E_{2})t} \braket{\varphi_{2} |\varphi_{1}}  + {\rm c.c.}  )  e^{(\Gamma_{1}+\Gamma_{2} ) t}  \, ]   \label{-ln}
		  \; .
	\end{aligned}
\end{equation}
For $H_{\pt}$ in PT-unbroken phase, $\Gamma_{n}=0$, $-\ln  {\rm Tr} \, \Omega(t) $ periodically oscillates; for  $H_{\pt}$ in PT-broken phase, some pairs of eigenvalues of it  become complex conjugate to each other, the biggest positive $\Gamma_{n}$ determines the dynamics of $-\ln  {\rm Tr} \, \Omega(t) $: it asymptotically decreases. For $H_{\apt}$ in PT-unbroken phase, $E_{n}=0$,  $\Gamma_{1, 2}$ determines the overall trend of $-\ln  {\rm Tr} \, \Omega(t) $:  it may be asymptotically decreasing (increasing or stable) without  oscillation; for  $H_{\apt}$ in PT-broken phase,   $\Gamma_{1, 2}$ determines the overall trend of $-\ln  {\rm Tr} \, \Omega(t) $:  it may be asymptotically decreasing (increasing or stable) with oscillation. 
% at least one of $E_{1, 2}$ is not zero.
Investigations \cite{kawabata2017information, wen2020observation} of information dynamics in (anti-) PT-symmetric systems show that phase transition in them is discontinuous, which is connected with the fact that Parity-Time (PT) symmetry  and anti-PT symmetry are discrete.
By Eq.(\ref{anyon}), we  know that the information dynamics of $H_{\varphi}$ is  the result of the interplay of  $H_{\pt}$ and $H_{\apt}$.  According to our analysis of $H_{\pt}$ and $H_{\apt}$, damped oscillation of  information dynamics is possible  for ${H}_{\varphi}$  in PT-unbroken phase or PT-broken phase,  showing that the phase transition in anyonic-PT symmetry is  continuous. The continuous  phase transition originates from the interplay of features of (anti-) PT symmetry and  the continuity of  anyonic-PT symmetry.

	\subsection*{Two-level systems}
	As  a proof-of-principle example, we consider generic two-level anyonic-PT symmetric system governed by  $H_{\varphi}$. With the parity operator P given by 
	$\begin{pmatrix} 0&1\\1&0\end{pmatrix}$, and the time reversal operator T being the operation of complex conjugation, $H_{\varphi}$   can be expressed as a family of matrices: 
	\begin{equation}
		H_{\varphi} = e^{-\ii \frac{\varphi }{2} }
		\begin{pmatrix}
			re^{\ii \theta }   & r_{1} e^{\ii\theta_{1}  }\\ r_{1} e^{-\ii\theta_{1}} & re^{-\ii\theta } 
		\end{pmatrix}, \label{varphi}
	\end{equation}
	where $ \varphi, \, r, \, \theta, \,  r_{1}, \,    \theta_{1}$  are   real.
	The energy eigenvalues of $H_{\varphi}$ are
	\begin{equation}
		E_{\pm } = e^{-\ii \frac{\varphi }{2} } (r\cos \theta \pm  \sqrt{\delta })  \; ,
	\end{equation}
	with
	\begin{equation}
		\delta =r_{1} ^{2} -r^{2} \sin ^{2} \theta \; .
	\end{equation}
	%In particular, when $r_{1}=r$, $\sqrt{\delta } = \left | r\cos \theta \right | $.
	%For  the PT  symmetric system,  PT symmetry is  unbroken for $\delta >0$  and spontaneously broken for $\delta <0$;   exceptional point is located at $\delta =0$ with the  order $n=2$.
When  $\delta >0$,  $H_{\varphi}$ in  PT-unbroken  phase;  when  $\delta <0$, $H_{\varphi}$ in  PT-broken phase;  the exceptional point of $H_{\varphi}$   locates  at $\delta =0$.
% with $\varphi=0$ for $H_{{\rm PT}}$  and  $\varphi=-\pi $ for $H_{{\rm APT}}$ and notice that $  \ii H_{{\rm PT}} =H_{\apt}$.   We then have	 we get Eq.(\ref{var}),
% \\&=e^{q \cdot 2tr\cos \theta}(   {\rm Tr}_{{\rm PT}}^{p} +  {\rm Tr}_{{\rm APT}}^{q}     -1    )
% \\
%&= \frac{1+b}{2}e^{q \cdot 2tr\cos \theta  }  \cos (2\ii p\sqrt{-\delta }t ) +  \frac{1-b}{2}e^{q \cdot 2tr\cos \theta  }    \cos (2q\sqrt{-\delta } t )
%	
%\begin{widetext} 
		When  $\delta > 0$, with  $a=\frac{r_{1}^2+r^2\sin ^{2}\theta }{\delta} \ge 1 $, 
	\begin{equation}
		\begin{aligned}
			{\rm Tr}\,\Omega(t) 
			&=e^{q \cdot 2tr\cos \theta  }  \cdot \frac{1-a}{2} \cos 2p\sqrt{\delta }t  \\ &+e^{q \cdot 2tr\cos \theta  } \cdot \frac{1+a}{2}\cos   2\ii q\sqrt{\delta }t    \; , 
				\end{aligned} 	\label{>0}
	\end{equation} 
	where $\frac{1-a}{2} \cos  2 p \sqrt{\delta }t $ is the feature of $H_{\pt} $ in PT-unbroken phase, and $\frac{1+a}{2}\cos   2 \ii q\sqrt{\delta }t  $ and $e^{q \cdot 2tr\cos \theta}$  are  the features of $H_{\apt} $ in PT-unbroken phase. 
	The interplay of $H_{\pt}$ and $H_{\apt}$ leads to  new novel properties unique to  $H_{\varphi}$.
	When $q \cdot 2r\cos \theta <0$,  the first term in Eq.(\ref{>0})  is  the equation of underdamped oscillation, with the undamped frequency $\omega^{2}=4(p^{2}\delta + q^{2}  r^{2}\cos ^{2} \theta)$, in particular, when $\left | r \right | =\left | r_{1}  \right | $,  $\omega^{2}=4  r^{2}\cos ^{2} \theta$;
	 the second term in Eq.(\ref{>0})  is the equation of overdamped oscillation,  with the undamped frequency $\omega^{2}=4q^{2}(r^{2}-r_{1}^{2})$,  in particular, when $\left | r \right | =\left | r_{1}  \right | $,  $\omega^{2}=0$.   So, Eq.(\ref{>0}) is a combination of the underdamped oscillation and the overdamped oscillation,  and  the undamped frequencies are independent of $\varphi$ when $\left | r \right | =\left | r_{1}  \right | $.
	When $q \cdot 2r\cos \theta >0$,   corresponding   amplified oscillations  can be analyzed in the same way. 
	When $\delta<0$,  with $b=\frac{r_{1}^2+r^2\sin ^{2}\theta }{-\delta} \ge 1 $,     we have 
	\begin{equation}
		\begin{aligned}
			{\rm Tr}\,\Omega(t) 
			&=e^{q \cdot 2tr\cos \theta  } \cdot  \frac{1+b}{2} \cos2 \ii p\sqrt{-\delta }t \\ & + e^{q \cdot 2tr\cos \theta  } \cdot \frac{1-b}{2} \cos 2q\sqrt{-\delta } t  \;  .
		\end{aligned}
		\label{<0}
	\end{equation}
So, similar to Eq.(\ref{>0}),  Eq.(\ref{<0})  is a combination of  underdamped oscillation and  overdamped oscillation and thus the information-dynamics patterns of $H_{\varphi}$ in PT-unbroken phase or PT-broken phase can be similar,   which shows that the  phase transition in the two-level anyonic-PT symmetric  is continuous. The asymptotically  stable  damped oscillation of $S_{1} (\Omega)$ of $H_{\varphi}$ in PT-broken phase is showed in FIG.(\ref{continuous}). 
%We show the continuous phase transition 
For $H_{\varphi_{1}}$ and $H_{\varphi_{2}}$ with $\varphi_{1}+\varphi_{2}=-2\pi $ or $2\pi$, the trace expressions of $H_{\varphi_{1}}$ and $H_{\varphi_{2}}$  are same. Thus, we only consider  $-\pi<\varphi<0$ ($p>0$, $q>0$). 
For  significantly   large $t$,
\begin{equation}
	{\rm Tr}\,\Omega(t) \sim  \frac{1+a}{4} e^{2q \lambda t      } 
	\label{varlim}
\end{equation}
Eq.(\ref{varlim}) determines the overall trend of Eq.(\ref{>0}), with $\lambda=r\cos \theta+\sqrt{\delta} $
($\lambda=0$ if and only if  $\left | r_{1}  \right | =\left | r \right | $ and $ r\cos \theta<0$ ).
There are  three   information-dynamics patterns for the   anyonic-PT symmetric systems: damped oscillation with an overall decrease (increase) and asymptotically  stable  damped oscillation,   as we show in FIG.(\ref{phe}). If we use the Hermitian quantum  Rényi entropy  or distinguishability, a three-fold degeneration and distortion happen, as we show in FIG.(\ref{compare}).  The three-fold degeneration and distortion happen in the PT-broken phase of $H_{\varphi}$ too, as we show in FIG.(\ref{continuous}) for the case of  asymptotically  stable  damped oscillation. 
The  degeneracy  is caused by the normalization of the non-unitary evolved density matrix $\Omega$, which washes out the effects of   decay parts $e^{\Gamma_{n}t} $ and thus 
leads to the loss of information  about the total probability flow between the open  system and the environment, while our approach based on the non-normalized density matrix reserves all the information related to the non-unitary  time evolution.  
The asymptotically  stable  damped oscillations and its relaxation time varying with $\varphi$ are showed in FIG.(\ref{fvarphi}).

	\section{Negative entropy  }
	
		\begin{figure}[t]
		\includegraphics[width=\linewidth]{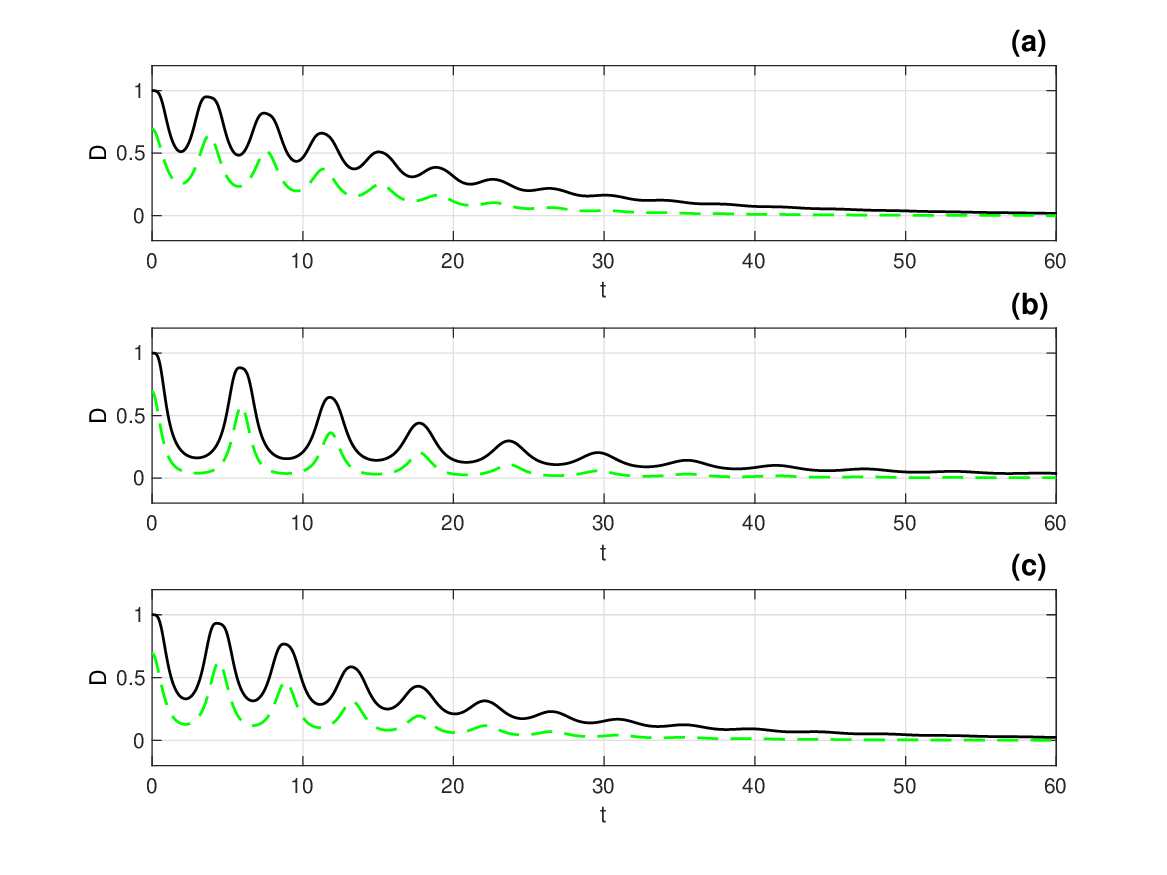}  %compare
		\caption{The black line represents distinguishability $D$, the dashed green line represents $S^H_{1}(\rho)$, i.e.,  von Neumann entropy.   $\varphi=-\pi/36$.
			(\textbf{a})  $r=0.8$, $\lambda>0 $, all parameters are same as FIG.(\ref{phe}a);           
			(\textbf{b})    $r=1.2$,  $\lambda<0 $, all parameters are same as FIG.(\ref{phe}c);
			(\textbf{c})  $r=1$,  $\lambda=0 $, all parameters are same as FIG.(\ref{phe}b). 
			While  $S_{1}(\Omega)$ and $-\ln {\rm Tr }(\Omega)$  show there are three   information-dynamics patterns for the   anyonic-PT symmetric systems: damped oscillation with an overall decrease (increase) and asymptotically  stable  damped oscillation. The  three patterns are  distorted by  $D$ and $S^H_{1}(\rho)$,  and degenerate to the same pattern as we show here.  We see that the distortion is related with the lower bounds of  $D$ and $S^H_{1}(\rho)$ being zero.
			\label{compare}     }
	\end{figure}

	Here comes the problem that $S_{\alpha}(\Omega)  $ can be negative
	and the comparison above in FIG.(\ref{compare}) gives a phenomenological justification for the necessity of it.  We go one step further and discuss  the negative entropy in NH open quantum system.       Entropy measures the degree of  uncertainty.  In the sense of classical statistical mixture, a closed  system with complete certainty  is possible, and thus  it's reasonable that the lower bound of von Neumann entropy  is  zero.
	 %i.e.,  the system in a pure state.  
	 However, an general open quantum system can't possess complete certainty since it constantly interacts with its external environment in a unpredictable way.  So, if we take the  entropy of closed  systems  as reference,   it's natural that for open quantum systems,  entropy  might be negative. 
	For example, unique properties of PT  symmetric systems are always  predicted and observed in classical or quantum  systems where gain and loss of energy or amplitude are balanced.  Then, we can reasonably expect that   different magnitudes of the balanced gain and loss  will  lead to  different  lower bounds of  entropy. %,    as we show in  fig.(\ref{4Mep}).  
	Negative entropy is possible and important in Hermitian physics too. 
	It is well known that
	quantum information theory has peculiar properties that cannot be found in its classical counterpart. For example, an observer’s uncertainty  about a system, if measured by von Neumann conditional entropy,  can become negative \cite{horodecki2005partial, rio2011thermodynamic, cerf1997negative}.  With  the density matrix of the combined system of $A$ and $B$ being  $\rho_{AB}$ (${\rm Tr} \,  \rho_{AB}=1   $),   von Neumann conditional entropy is defined as
	\begin{equation}
		S(A|B)=-{\rm Tr}(\rho_{AB}\ln \rho_{A|B}  )
		\label{neg}
	\end{equation}
which is based on a conditional ``amplitude" operator $\rho_{A|B}$ \cite{cerf1997negative}.  The eigenvalues of $\rho_{A|B}$ can exceed 1 and  it is precisely for 
this reason that the von Neumann conditional entropy can be negative \cite{cerf1997negative}.  
For our purpose, the similarity between Eq.(\ref{neg}) and Eq.(\ref{s1}) inspires a  comparison of  the role of non-normalized density matrix $\Omega$ in NH entropy $S_{1}(\Omega)$ and the role of $\rho_{A|B}$ in  von Neumann conditional entropy $S(A|B)$,  we remark that  the mathematical reason why $S_{1}(\Omega)$  can be negative is similar to $S(A|B)$, as ${\rm Tr}\,\Omega $ can exceed 1.  
The strong correlation between  $-\ln {\rm Tr}\,\Omega $ and $S_{1}(\Omega)$ also suggests that ${\rm Tr}\,\Omega > 1$ will lead to negative entropy. Negative von Neumann conditional entropy has been given  a physical interpretation in terms of how much quantum communication is needed to gain complete quantum information \cite{horodecki2005partial}. Furthermore, a direct thermodynamical interpretation of negative conditional entropy is  given in \cite{rio2011thermodynamic}. For NH entropy, our results above  demonstrate that allowing  NH entropy to be negative is necessary and inevitable if we want to characterize the information dynamics of NH system properly. %, and the non-normalized density matrix $\Omega$ is essential. 

\paragraph*{Conclusion and outlook.\,---}
	%While Parity-Time (PT) symmetry  and anti-PT symmetry are discrete, anyonic-PT symmetry as their complex generalization is continuous respect to the phase parameter in a way similar to rotation symmetry. 
	We investigate the non-Hermitian (NH) quantum Rényi entropy dynamics of anyonic-PT symmetric  systems through a new information-dynamics description $-\ln {\rm Tr}\,\Omega $, which is found to be  synchronous and correlated with NH quantum Rényi entropy.  Our results show: in contrast to the discontinuous phase transition in  (anti-) PT-symmetric  systems,   the phase transition in anyonic-PT symmetry is  continuous. The continuous  phase transition originates from the interplay of features of (anti-) PT symmetry and  the continuity of  anyonic-PT symmetry.  We find  there are  three   information-dynamics patterns for  anyonic-PT symmetric systems: damped oscillation with an overall decrease (increase) and asymptotically  stable damped oscillation,   which are three-fold degenerate and distorted if we use the Hermitian quantum  Rényi entropy  or distinguishability. 
	The  discussion of  the degeneracy and distortion   serves as  a  justification for  negative NH quantum Rényi entropy. We further  explore the mathematical reason and physical meaning of  the negative entropy in open quantum systems,  revealing a  connection between negative NH entropy and negative quantum conditional entropy as both quantities can be negative for similar mathematical reasons.   
	Since the physical interpretation and the following applications   of negative quantum conditional entropy  are  successful and  promising  \cite{horodecki2005partial, rio2011thermodynamic, cerf1997negative}, our work opens up the new  journey of rigorously investigating  the  physical interpretations and the application  prospects of negative entropy in open quantum  system.

	\paragraph*{Acknowledgements.\,---}
	This work was supported by the National Natural Science Foundation of China (Grant Nos. 12175002, 11705004, 12381240288), the Natural Science Foundation of Beijing (Grant No. 1222020), the Project of Cultivation for Young top-notch Talents of Beijing Municipal Institutions (BPHR202203034). 
	Zhihang Liu acknowledges valuable discussions with  Daili Li.

	\bibliography{refer.bib}
	
	\begin{widetext}
		
	\section*{Appendix}

	\setcounter{equation}{0}

\renewcommand\theequation{A.\arabic{equation}} %（A1）

	\subsection*{Derivation of Eq.(\ref{>0}) and Eq.(\ref{<0})  }
	
	Define the two-level PT-symmetric  $H_{\pt} $ as 
		\begin{equation}
		H_{\pt} = 
		\begin{pmatrix}
			re^{\ii \theta }   & r_{1} e^{\ii\theta_{1}  }\\ r_{1} e^{-\ii\theta_{1}} & re^{-\ii\theta } 
		\end{pmatrix}, 
	\end{equation}
	and decompose it in Pauli matrix, 
	\begin{equation*}
		H_{{\rm PT}}=r\cos \theta I+r_{1} \cos \theta _{1} \sigma _{1} -r_{1} \sin  \theta _{1} \sigma _{2}  +\ii r\sin \theta \sigma _{3} \label{pauli}  \; .
	\end{equation*}
	 Define
	\begin{gather}
		M=r_{1} \cos \theta _{1} \sigma _{1} -r_{1} \sin  \theta _{1} \sigma _{2}  +\ii r\sin \theta \sigma _{3} 
		=\begin{pmatrix}
			\ii r\sin \theta     & r_{1} e^{\ii \theta_{1}  }\\
			r_{1} e^{-\ii \theta_{1}} & -\ii r\sin \theta
		\end{pmatrix} .
	\end{gather}
	 It is easy to verify that
	\begin{equation}
		M^{2} =\delta I   , 
	\end{equation}
	with $	\delta =r_{1} ^{2} -r^{2} \sin ^{2} \theta  $.
	The time-evolution operator  $U_{\varphi }$ of  $H_{\varphi}= e^{-\ii \frac{\varphi }{2} } H_{{\rm PT}} $ is %$H_{\varphi }$
	\begin{equation}
		\begin{aligned}
			%\begin{split}
			U_{\varphi } 
			&=e^{-\ii tH_{\varphi } } \\
			&=e^{-\ii te^{-\ii \frac{\varphi }{2} } r\cos \theta  }\cdot
			(\sum_{k=0 }^{\infty} \frac{(-\ii te^{-i\frac{\varphi }{2} }  )^{2k}\cdot M^{2k}  }{2k!}  
			+\sum_{k=0 }^{\infty} \frac{(-\ii te^{-i\frac{\varphi }{2} }  )^{2k+1}\cdot M^{2k+1}  }{(2k+1)!} )\\
			\label{e5}
			%\end{split}
		\end{aligned}.
	\end{equation}
	When $\delta =0$,  $M^{2} =\delta I =0$, we have
	\begin{equation}
		U_{\varphi } 
		=e^{-\ii te^{-\ii \frac{\varphi }{2} } r\cos \theta  }\cdot(-\ii te^{-\ii \frac{\varphi }{2} } M +I)
		\label{e6}.
	\end{equation}
	When $\delta \ne 0$, %we can go further and get the expression
	\begin{equation}
		\begin{aligned}
			U_{\varphi } 
			&=e^{-\ii te^{-\ii \frac{\varphi }{2} } r\cos \theta  }\cdot	(\sum_{k=0 }^{\infty} \frac{(-\ii te^{-\ii \frac{\varphi }{2} }   )^{2k}\cdot (\sqrt{\delta } )^{2k}I  }{2k!}  +\sum_{k=0 }^{\infty} \frac{(-\ii te^{-\ii \frac{\varphi }{2} }  )^{2k+1}\cdot (\sqrt{\delta})^{2k+1} \frac{M}{\sqrt{\delta } }  }{(2k+1)!} )\\
			&=e^{-\ii te^{-\ii \frac{\varphi }{2} } r\cos \theta  }   \cdot	(    \cos (te^{-\ii \frac{\varphi }{2} } \sqrt{\delta }  ) I    -       \ii \frac{\sin (te^{-\ii \frac{\varphi }{2} } \sqrt{\delta }  )}{\sqrt{\delta }}M     ) .
			\label{e10}
		\end{aligned}
	\end{equation}
	Denote
	\begin{equation}
		M_{1}=  \cos (te^{-\ii \frac{\varphi }{2} } \sqrt{\delta }  ) I    -       \ii \frac{\sin (te^{-\ii \frac{\varphi }{2} } \sqrt{\delta }  )}{\sqrt{\delta }}M.  
	\end{equation}   
	With $	\Omega (0)=\frac{1}{2} I  $,	
	\begin{equation}
			\begin{aligned}
		\Omega (t) &=U_{\varphi }	\Omega (0) U_{\varphi }^{\dagger } \\
		&=\frac{1}{2}e^{q \cdot 2tr\cos \theta   }  \cdot M_{1}	M_{1}^{\dagger} \; .  
	\end{aligned}  
\label{a8}	\end{equation}
	When  $\delta > 0$, with  $a=\frac{r_{1}^2+r^2\sin ^{2}\theta }{\delta} \ge 1 $, we get Eq.(\ref{>0}),
	\begin{equation}
		\begin{aligned}
			{\rm Tr}\,\Omega(t) 
			=e^{q \cdot 2tr\cos \theta  }  \cdot \frac{1-a}{2} \cos 2p\sqrt{\delta }t  +e^{q \cdot 2tr\cos \theta  } \cdot \frac{1+a}{2}\cos   2\ii q\sqrt{\delta }t    \; .
		\end{aligned} 	
\label{a9}	\end{equation} 
	When $\delta<0$,  with $b=\frac{r_{1}^2+r^2\sin ^{2}\theta }{-\delta} \ge 1 $,     we get Eq.(\ref{<0}),
\begin{equation}
	\begin{aligned}
		{\rm Tr}\,\Omega(t) 
		=e^{q \cdot 2tr\cos \theta  } \cdot  \frac{1+b}{2} \cos2 \ii p\sqrt{-\delta }t + e^{q \cdot 2tr\cos \theta  } \cdot \frac{1-b}{2} \cos 2q\sqrt{-\delta } t  \;  .
	\end{aligned}
\label{a10} \end{equation}
	%So, similar to Eq.(\ref{var0}),  Eq.(\ref{<0})  is a combination of  underdamped oscillation and  overdamped oscillation. All the trace expressions can be derived in the same way. 	\subsection*{Consequence of   normalization }	
By Eq.(\ref{a8}), 	Eq.(\ref{a9}) and  Eq.(\ref{a10}),  we know that  the normalization procedure $	\rho(t)=\frac{	\Omega (t)}    {{\rm Tr}\,\Omega(t)}$ washes out the decay part $e^{q \cdot 2tr\cos \theta}$ ($q=1$ for $H_{{\rm APT}} $, $q=0$ for $H_{{\rm PT}} $) and causes loss of information about the total probability flow between the NH open  quantum system and the environment.
%	\begin{equation}
%		\begin{aligned}
%			\rho(t)&=\frac{	\Omega (t)}    {{\rm Tr}\,\Omega(t)} \\
%			&=\frac{\frac{1}{2} M_{1}	M_{1}^{\dagger}}{ {\rm Tr}_{{\rm PT}}^{p} +  {\rm Tr}_{{\rm APT}}^{q}     -1   }
%	\; .	\end{aligned}
%	\end{equation}

\end{widetext}

\end{document}